David PREVOST
Sylvain LAVERNHE
Claire LARTIGUE

# FEED DRIVE SIMULATION FOR THE PREDICTION OF THE TOOL PATH FOLLOW UP IN HIGH SPEED MACHINING

This paper deals with an advanced modeling of the feed drives of a five axis machine tool within the context of High Speed Machining. The management of the multi axes as well as high velocities causes problems to the set machine tool – Numerical Controller throughout the trajectory execution process. As a result, many errors are introduced during machining all process long affecting the surface quality. The paper aims at modeling the feed drive dynamics during trajectory follow-up including the current, the velocity and position loops as well as the feed forward terms, which characterize classical drives on actual HSM machines. It concerns translational axes as well as rotary axes. A procedure of identification is implemented. Performances of the model are assessed by the comparison between simulated tool paths to the real one. Experimental verifications of the virtual axis model are detailed for three and five axis trajectories presenting various types of geometrical discontinuities.

## 1. INTRODUCTION

The elaboration process of free-form parts using 5-axis machining is a complex process which involves several steps. Based on geometrical and functional specifications, a CAD model of the part to be machined is created. From this CAD model, the trajectory of the tool which allows the complete part machining is calculated during the CAM step. The trajectory, generally calculated as a set of points and corresponding federates, defines the CL-file (or apt file) which is transmitted to the Numerical Unit (NC) of the machine tool. The NC unit translates the CAM trajectory into a set of axis commands to realize the part machining. The last step is the actual machining leading to the machined part. The process thus involves a large number of parameters, data transformations and information exchanges, which may affect global productivity and final part quality [16].

In order to apprehend the control of quality, the process can be split into 3 major stages, each one being considered separately (Fig. 1). The first one is the CAM stage which transforms the CAD model into a tool trajectory giving the CL-file. Errors associated to this stage are generally linked to the discretization parameters used to calculate the tool path [3]. The second stage concerns the transformation of the trajectory into axis commands by the NC unit. Two main treatment levels are associated to this stage [4]. The level 1 (or numerical level) interprets in real time the program lines of the CL-file to develop axis commands. The level 2 (or analogue/digital level) corresponds to axis cards and drives which realize, after the converting of analogue data into digital data, the axis controls.

The third stage is the actual machining on the machine tool, which in turn can be split into two levels. The first one, corresponding to the mechanical chain, creates the relative movement tool/part from the axis commands previously calculated. The second level, which is also the last level of the whole process, consists in the physical cutting process. Errors or loss of geometrical quality associated to this level come from both thermals and mechanical phenomena due to the cutting process as well as the structure deformations and dynamics phenomena occurring during machining.

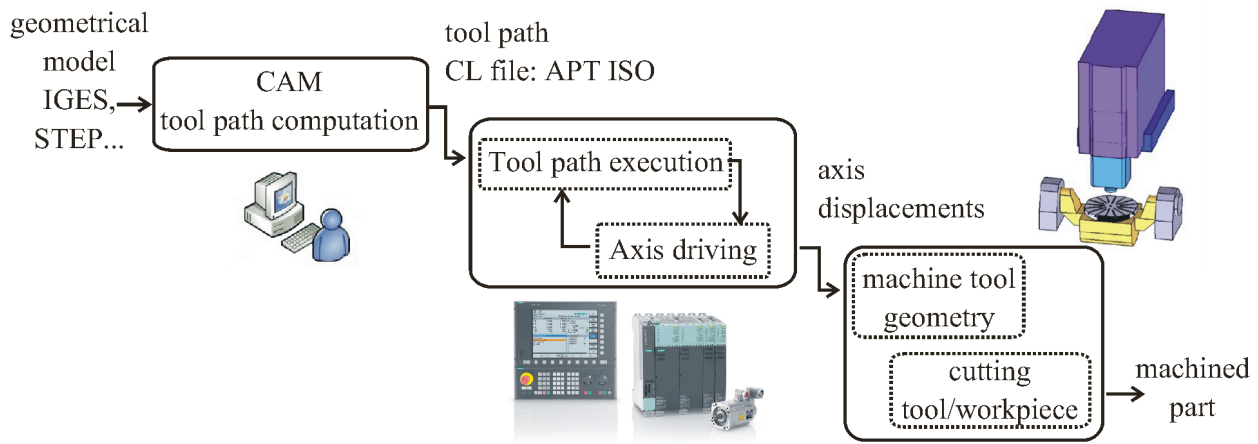

Fig. 1. High Speed Machining process

The prediction of such errors would require the modelling of the structure of the machine tool, the modelling of the dynamics during machining and the modelling of the mechanics of the cutting process itself, including tool and part deformations.

The present paper more particularly deals with errors and gaps associated to the NC treatments and the kinematics. For instance, interpolation operations or the Inverse Kinematics Transformation (IKT) calculated by the NC as well as performance of the servo drives and the kinematics directly affect the part geometry. With the objective of error prediction, a first model of trajectory execution has been defined [10], [11], [12]. By integrating kinematics performances, this model allows the simulation of the actual tool follow-up during machining, both at the kinematical level of each axis by determination of position, velocity, acceleration and jerk and at the relative movement tool/part. Results have highlighted that the NC treatment, in particular the stage of interpolation (level 1) is source of errors and approximations leading to geometrical deviations on the machined parts. However, in order to improve the simulation as regards the real behaviour, a modelling of the feed drives (level 2) separated from the modelling of the interpolator is necessary. It would thus be possible to separate errors coming from the interpolator from those coming from the feed drives [15]. Therefore, the paper aims at modelling the NC drives with the objective of evaluating errors imputable to level 2. Simulation results must be quite precise to be compared with all the errors involved. The axis drive model proposed is deliberatively simple to apply to any type of axis (translation or rotation) and whatever the industrial high-speed machine tool.

# 2. TOOL PATH FOLLOW-UP MODELING IN SERVO SYSTEMS

## 2.1. MACHINE TOOL AXIS DRIVE PRINCIPLE

Common feedback structures used in industrial CNC drives are often based on a cascade control structure. A cascade structure consists of several nested loops which have increased dynamics (Fig. 2). The current inner loop aims at controlling the motor torque as the torque is directly proportional with the armature current. Its high bandwidth prevents from disturbances. The current loop is closed using a proportional integral (PI) control. The velocity central loop is also closed using a PI controller. It presents a smaller bandwidth controller. The position loop is the most external one with a larger sampling period. This last loop is closed using to a proportional gain. A velocity and torque feedforward actions are added to reduce the tracking error resulting from the lack of integrative action in the position controller [18].

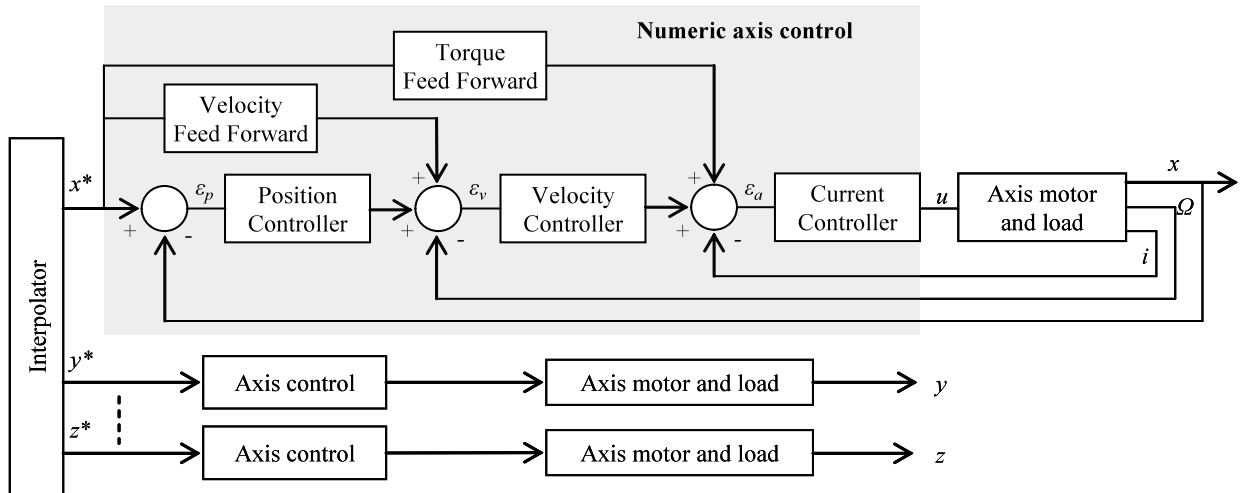

Fig. 2. Axis control structure

The structure defined above is commonly used in industrial CNC drives with the major drawback that the structure is not open, essentially for historical and robustness reasons.

## 2.2. DYNAMIC MODEL OF FEED DRIVES

The modelling of tool path follow-up proposed in the paper relies on the feed drive structure presented in Fig. 2. In the next, axes are modelled in an independent way, neglecting the existing coupling interactions. To illustrate our purpose, only the modelling of the X translation axis is detailed here. Obviously, a similar modelling is built for the other translation axes as well as for the rotational axes. Axes affected by gravity are modelled quite differently: gravity is taken into account by adding a resistant torque to the motor. For a vertical axis, the torque is constant; for a tilting table the torque depends on the angular axis position.

The modelling concerns the level 2 of the CNC treatment and the realization of the movements by the kinematical chain. For each axis, the input of the model is the command calculated by the interpolator and transmitted to the axis cards. This command is the Position Set point at Entrance of the Controller (PSEC) [2]. The main output of the model is the simulated position (SP). Other inner variables of the model such as current, motor torque, velocity and acceleration axis can be collected during simulation.

As the model must be generic and simple, only few physical phenomena are deliberately modelled. Dynamic behaviour, such as vibrations, stiffness in joints, backlashes and other filters or compensations are not modelled at this step [17], [1], [20], [19]. Nevertheless, to account for these phenomena and other numerical treatments, an adjustment parameter, called "delay" is set in both the position and feedforward loops. The complete model is implemented using the function Simulink of Matlab© (See Appendix).

In the next section, a strategy of parameter identification of the model is detailed in order to determine the axis dynamics.

## 3. MODEL PARAMETER IDENTIFICATION

More generally, all the feedback controller parameters are available in the CNC, as well as the sampling periods of each loop. The parameters to be identified are thus, the equivalent inertia, the friction coefficients and the feedforward compensations.

In literature, many techniques for parameter identification exist. Some of them rely on frequency analysis by disconnecting the servo loops [21], [5]. Global techniques are also proposed so that the overall closed loop dynamics is determined by running a standardized G-code and capturing the commanded and measured positions [6]. Indeed, time-dependant variables, such as current, position, velocity and acceleration can be captured on the fly in most modern CNC systems. This method is used to determine parameters of the servo axis model.

### 3.1. FRICTION MODELING AND PARAMETER IDENTIFICATION

In axis control, the dominant source of disturbance is friction. Several studies lead to the modelling of friction through relationships between friction, position, speed and temperature of rigid bodies [7], [14]. In this paper, the choice of a friction model depending only on axis velocity is made, as suggested in [14], [13]. Static model considered here is the combination of viscous friction and Coulomb friction. The relevance of this modelling will be assessed during the identification phase.

The modelling of the friction law is obtained from the dynamic moment equation projected onto the axis of the motor shaft. This yields to Eq. 1., where $C_m$ stands for the motor torque, $C_r$ for resistant torques, $J_{eq}$ for the equivalent inertia and $\Omega_m$ for the angular velocity of the motor.

$$C_m - C_r = J_{eq} \cdot \frac{d\Omega_m}{dt} \qquad (1)$$

When acceleration is null (constant velocity displacement), the motor torque, which is directly proportional to the current, is equal to the resistant torque. It is thus possible to determine the real friction law by successive measures of the current for different constant velocity displacements. Then, a trend curve applied to experimental point models the real behaviour of friction with enough accuracy (Fig. 3).

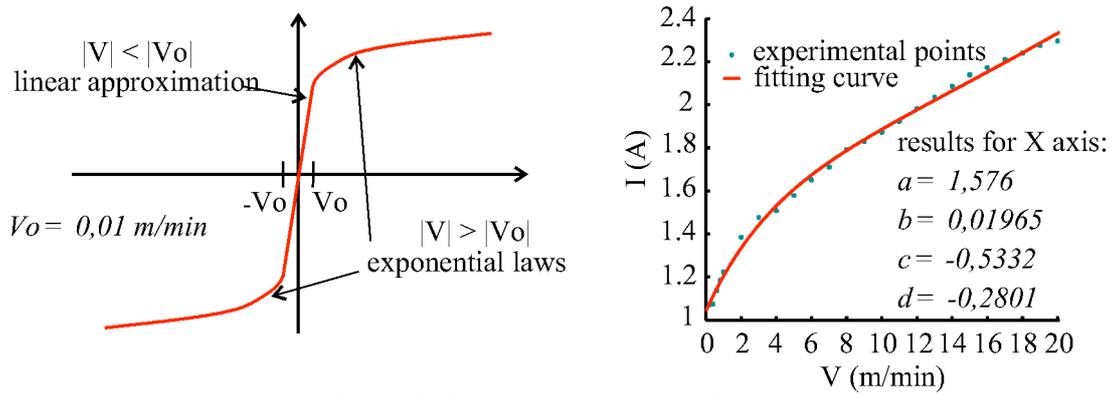

Fig. 3. Friction law and experimental results

Concerning the motor parameters, electric parameters are provided by the machine-tool builder. Unfortunately, mechanical parameters which account for embedded masses and other inertias also require an experimental procedure of identification.
Equivalent inertia related to the motor shaft is calculated from a series of current measurements with non null acceleration displacements according to Eq. 2 which results from Eq. 1 where $K_t$ stands for the own torque constant of the motor.

$$J_{eq} = \frac{K_t \cdot i - C_r}{\frac{d\Omega_m}{dt}} \qquad (2)$$

### 3.2. FEEDFORWARD AND SHIFT PARAMETERS

Feedforward is a common technique used to cancel dragging differences on the position [9]. Two types of feedforward are modelled: the torque feedforward and the velocity feedforward which can be disabled, used separately or together. Feedforward set points (VFFWS and TFFWS) are generally proportional to velocity and acceleration set points at the entrance of the controller. Therefore, constants associated to the feedforward compensations can be evaluated from the measurement of the feedforward set points.
Following the identification stage, the step of parameter adjustment is performed.

## 3.3. MODEL ASSESSMENT

In order to assess the model, comparisons between simulations and experimental measures of the real tool position are carried out on a five axis milling centre Mikron UCP 710 HSM – (RRTTT machine structure) – equipped with a Siemens Numerical Controller Sinumerik 840D. Different types of trajectories are tested, involving one, two or three axes. The measured signals for each type of trajectory are the position set points (mm), the current (A) and the velocity (m/min).
For all the tests, the machine is working without any charge and firstly the feedforward actions are cancelled. Testing one-axis displacement only permits to check independently each axis while being free from coupling problems. This series of test concerns both the axes of translation and the axes of rotation. Whatever the axis, all the experiments lead to similar results. Therefore, only tests of the X axis of translation are reported in Fig. 4. A linear displacement is programmed with a velocity of 15 m/min., with a jerk limited acceleration profile. The deviations between the measured and the simulated position oscillate around zero but do not exceed 4 µm except during critical phases of acceleration or deceleration. In these cases, differences can reach 6 or 8 µm. Considering that similar remarks can be made for other axes, differences between simulation and actual values are weak enough to assess the model of each axis independently, with or without feedforward actions.

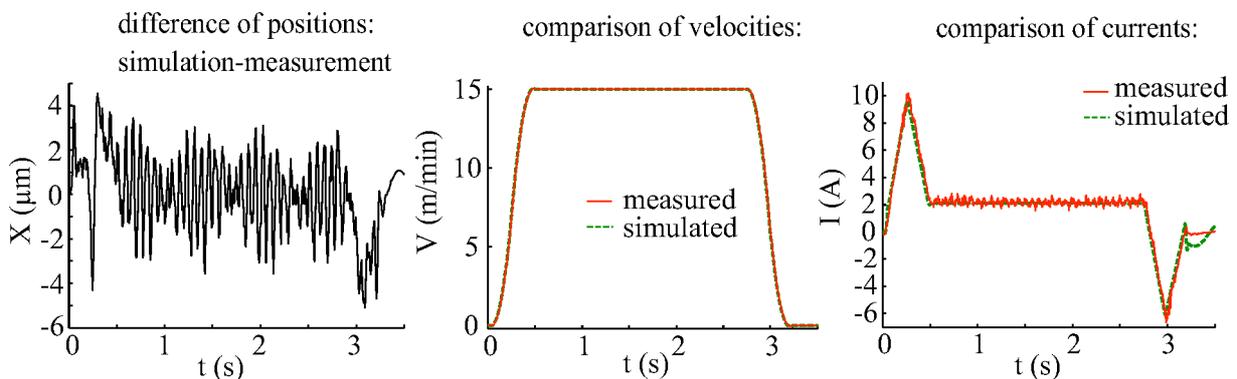

Fig. 4. Simple X translation test

Other tests conducted for more complicated trajectories lead to the same results: simulated trajectories match the measured ones although very local gaps between simulation and measure exist.
As a conclusion of the assessment step, the simulated follow-up gives a good representation of the real one. Moreover, the precision of the simulations seems to be good enough to predict errors or deviations resulting from the level 2, and their impact onto the part geometry.

# 4. APPLICATIONS TO MACHINED PARTS

This section aims at evaluating geometrical errors of the machined parts due to the trajectory follow-up during high-speed machining by simulation. For this purpose, the model presented above is used in the simulation of the machined part geometry. In parallel, the actual machining of the part is carried out on the machine tool Mikron UCP710. The actual geometry is measured using a confocal z-axis extended field sensor (STIL®) to be compared with the simulated one. Details of the procedure are given in Fig. 5.

The inputs of the axis follow-up model are the axis position set points calculated by the interpolator of the CNC. Simulated axis positions, resulting from the follow-up simulation as detailed in the previous section, are used to build the simulated tool trajectory by means of the direct kinematics transformation. From this virtual trajectory, the geometry of the part is simulated using a Nbuffer material removing model [8]. Result of simulation is then compared to the actual part.

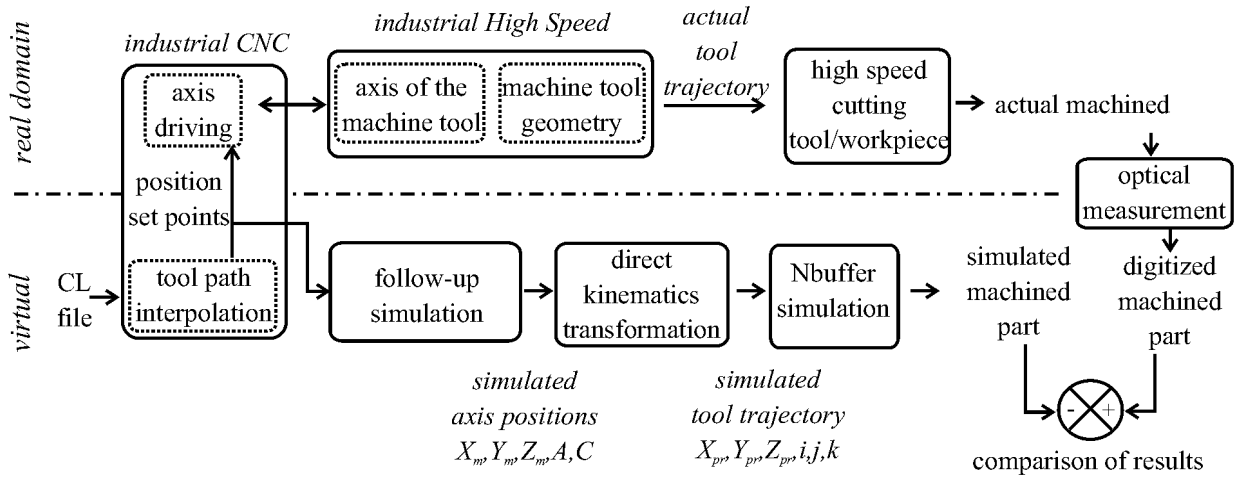

Fig. 5. Simulation of the machining process

To bring out the influence of high velocities, two types of tool path are superposed. First, a reference is defined by the machining of the part with a low-programmed feedrate. Deviations caused by follow-up can be neglected. Then, the same tool path is executed with a high velocity.

Different tests parts characteristic of moulds and dies machining are used (Fig. 8). The first one illustrates a case of point milling with both the X and Z axes. The second one corresponds to the flank milling of a sharp corner using the X and Y axes. The last case involves the 5 axes, with in particular the rotational axes A and C.

The first test-part consists of two planes connected in tangency by a portion of cylinder. The machining strategy is parallel planes in one way (XZ planes), with constant distance between planes of 0.25mm. The diameter of the ball-end mill tool is 10mm. The low programmed feedrate is set to 0.1m/min whereas the high feedrate is 10m/min. Only one path is executed with the high feedrate. Fig. 6 compares the measured geometry obtained after machining (left picture) with the simulated geometry obtained by virtual machining (right picture). Results focus on the connection zone.

A mark is clearly visible for the measured part as well as for the simulated one. The maximal depth is around 0.02mm for both cases. Considering that cutting conditions are consistent with the material (aluminium alloy) and the thickness used for finishing (0.05mm), the source of deviations is probably linked to the trajectory follow-up.

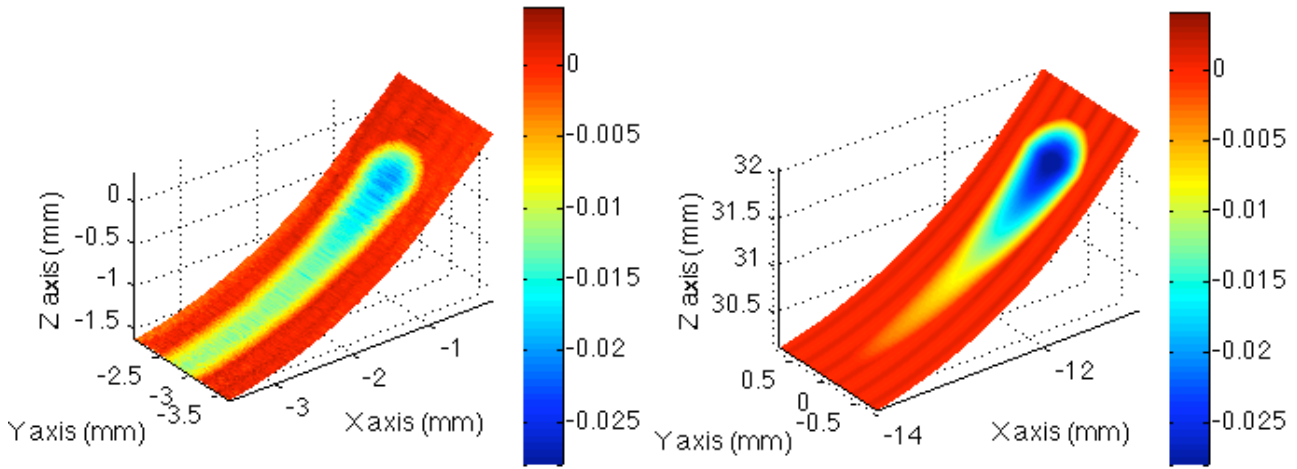

Fig. 6. Results of the surface geometry for the first test part

The second test part represents the finishing by flank milling of a vertical corner with a cylindrical tool which diameter is 10mm. The reference is machined in one tool path with an axial depth of cut of 10mm and a low federate. Then the same tool path is translated of about 5mm along the Z axis, and is re executed with the high feedrate; the axial depth of cut is thus of 5mm. Fig. 7 compares the measured geometry to the simulated one.

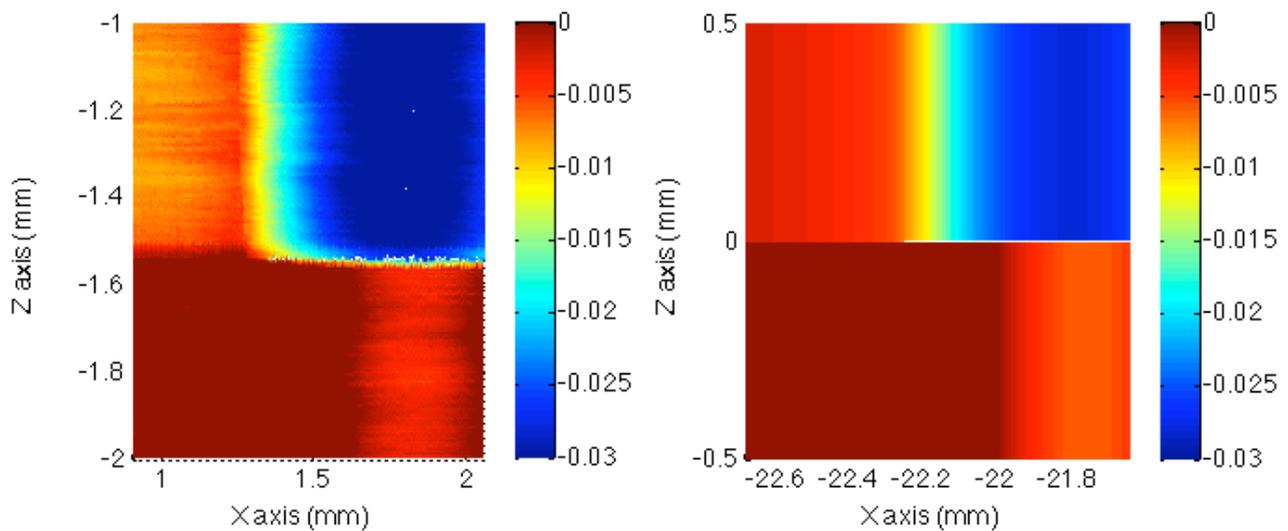

Fig. 7. Results of surface geometry for the second test part.

Simulated and measured values are quite similar. With low feedrate, a little mark is visible (0.005mm). With high feedrate, maximal geometrical error on the surface is around 0.03mm. As for the previous case, this example highlights that trajectory follow-up at high speed causes deviation, in particular when machining trajectory discontinuities. Remaining differences between simulations and measures are supposed to be due to the real geometry of the machine tool, actual tool geometry and cutting. Nevertheless, it can also be brought out that the virtual machining is a good prediction of the actual machining.

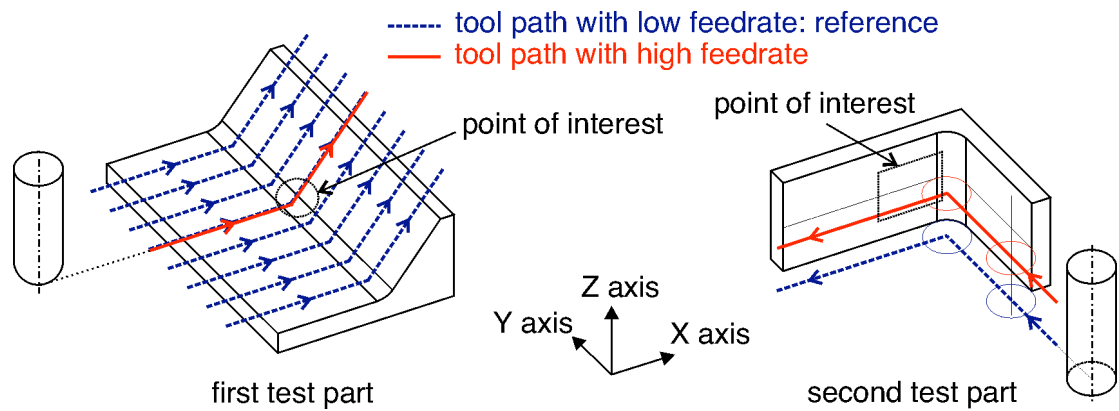

Fig. 8. Geometry of test parts n°1 and 2.

For the last case, the trajectory is a 5-axis trajectory which involves the rotational axes with large displacements. For this example, we only consider geometrical errors on the part caused by the follow-up of rotational axes.
Although programmed feedrates are important, solicitations on the rotational axes are not sufficient to alter the trajectory follow-up and to generate large marks on the part. Deviations measured on axis during machining and simulated by the model are around 0.001 degrees. This kinematical behaviour is due to low acceleration and jerk performances of the rotary and the tilting table of the HSM centre used for the experiments.

## 5. CONCLUSION

Within the context of HSM, it is generally difficult to separate the sources of errors linked to each stage of the machining process. Therefore, an interesting issue concerns the complete modeling of an industrial HSM machine so that each source of geometrical errors could be evaluated. Starting from previous work, a model of the trajectory execution performed by the NC unit is proposed. The interest of the simulation is to predict how large and where these deviations appear. In the paper, only the deviations due to the servo motion and the trajectory follow-up are considered. A simulation model is proposed to evaluate the importance of these deviations and is implemented in a structure of virtual machining allowing the prediction of the machined geometry. Through various cases, the importance of the follow-up deviation is investigated. If it can not be neglected when machining geometrical discontinuities in 3-axis milling, its importance is less in the case of 5 axes.

Indeed, in such a case, as the rotational axes are less dynamic, the follow-up is less altered. Future work will focus on the integration of the virtual machining chain.

ACKNOWLEDGEMENTS

This work was carried out within the Manufacturing 21 working group, which comprises 18 French research laboratories.
The topics approached are:
- the modelling of the manufacturing process;
- the virtual machining;
- the emergence of new manufacturing methods.

REFERENCES

[1] BARRE P. J., *Modélisation du comportement dynamique et commande d'une machine-outil agile*, Mécanique & Industries, 2002, Vol. 3.
[2] BLOCH S., DENEUVILLE E., TAN L., *Innovative feed rate optimisation technique*, 3$^{rd}$ international conference on metal cutting and High Speed Machining, 2001.
[3] DUC E., LARTIGUE C., TOURNIER C., BOURDET P., A new concept for the design and the manufacturing of free-form surfaces: The machining surface, Annals of the CIRP, 1999, Vol. 48/1, pp 103-106.
[4] DUGAS A., *Simulation d'usinages de formes complexes*, PhD thesis, Ecole Centrale de Nantes, Université de Nantes, 2002.
[5] ERKORKMAZ K., ALTINTAS Y., *High speed CNC system design. Part II: modelling and identification of feed drives*, International Journal of Machine Tools and Manufacture, 2001, Vol. 41, pp. 1487-1509.
[6] ERKORKMAZ K., WONG W., *Rapid identification technique for virtual CNC drives*, International Journal of Machine Tools and Manufacture, Vol. 47, pp. 1381-1392, 2007.
[7] KIKUUWE R., TAKESUE N., SANO A., MOCHIYAMA H., FUJIMOTO H., *Fixed-step friction simulation: from classical Coulomb model to modern continuous models*, International Conference on Intelligent Robots and Systems, Canada, 2005.
[8] JERARD R.B., DRYSDALE R.L., HAUCK K., SCHAUDT B., MAGEWICK J., *Sculptured Surfaces - Methods for Detecting Errors in Numerically Controlled Machining of Sculptured Surfaces*, IEEE Computer Graphics & Applications, 1989, pp. 26-39.
[9] LAMBRECHTS P., BOERLAGE M., STEINBUCH M., *Trajectory planning and feedforward design for electromechanical motion systems,* Control Engineering Practice, 2005, Vol. 13, pp. 145-447.
[10] LAVERNHE S., *Prise en compte des contraintes associés au couple MO-CN en génération de trajectoire 5 axes UGV*, PhD thesis, Ecole Normale Supérieure de Cachan, 2006.
[11] LAVERNHE S., QUINSAT Y., TOURNIER C., LARTIGUE C., MAYER R., *NC-simulation for the prediction of surface finish in 5-axis High Speed Machining,* Proceedings of 3rd International Conference High Performance Cutting (HPC), Dublin, Irlande, 2008, Vol. N° 1, pp. 387-396, pp. 12-13.
[12] LAVERNHE S., TOURNIER C., LARTIGUE C., *Optimization of 5-axis high-speed machining using a surface based approach,* Computer Aided Design, 2008, Vol. N° 40, pp. 1015-1023.
[13] MENON K., KRISHNAMURTHY K., *Control of low velocity friction and gear backlash in a machine tool feed drive system Mechatronics*, 1999, Vol. 9, pp. 33-52.
[14] OLSSON H., ASTRÖM K.J., CANUDAS DE WIT C., GÄFVERT M., LISCHINSKY P., *Friction models and friction compensation*, Lund University, 1997.

[15] RAMESH R., MANNAN M.A., POO A.N., *Tracking and contour error control in CNC servo systems*, International Journal of Machine Tools and Manufacture, 2005, Vol. 45, pp. 301-326.
[16] SCHMITZ T. L., ZIEGERT J. C., CANNING J. S., ZAPATA R., *Case study: A comparison of error sources in high-speed milling*, Precision Engineering, 2008, Vol. N° 32, pp. 126-133.
[17] SIEMENS, *Description of functions* – Sinumerik 840D/840Di/810D, www.automation.siemens.com/doconweb, 2002.
[18] SUSANU M., DUMUR D. LARTIGUE C., TOURNIER C., *Improving performance of machine tools with predictive axis controllers within an open architecture Framework*, 3rd International Conference on Advanced Manufacturing Technology, Kuala Lumpur, Malaysia, 2004
[19] WHALLEY R., EBRAHIMI M., ABDUL-AMEER A.A., *Hybrid modelling of machine tool axis drives*, International Journal of Machine Tools and Manufacture, 2005, Vol. 45, pp. 1560-1576.
[20] YEUNG C.-H., ALTINTAS Y., ERKORKMAZ K., *Virtual CNC system. Part I. System architecture*, International Journal of Machine Tools and Manufacture, 2006, Vol. 46, Chap. 10, pp. 1107-1123.
[21] YEUNG C.-H. , ALTINTAS Y., ERKORKMAZ K., *Virtual CNC system. Part II. System architecture*, International Journal of Machine Tools and Manufacture, 2006, Vol. 46, Chap. 10, pp. 1124-1138.

# Appendix

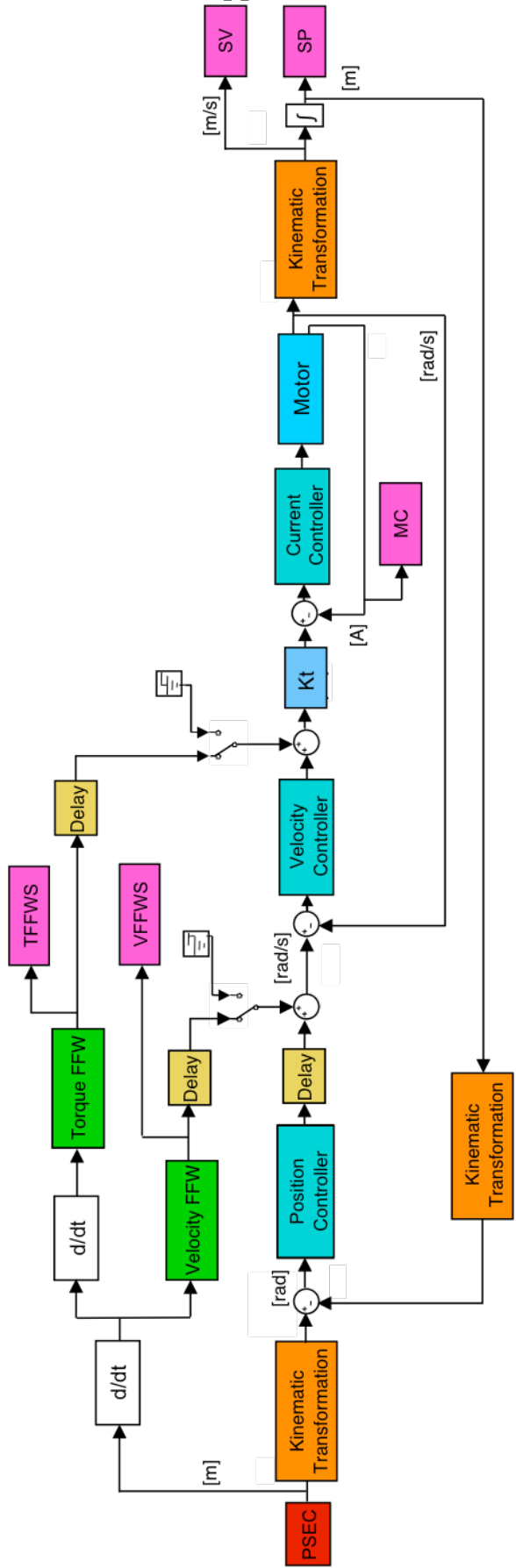